\documentclass[aps,prev,twocolumn,preprintnumbers,floatfix,nofootinbib]{revtex4-1}
\pdfoutput=1
\usepackage{amssymb,graphicx}
\usepackage{epstopdf}
\usepackage{mathrsfs}
\usepackage{amssymb}
\usepackage{verbatim}
\usepackage{xcolor}
\usepackage{multirow}
\usepackage{amsmath}
\usepackage{braket}
\usepackage{subfig}
\usepackage{slashed} 
\usepackage{float}
\usepackage[normalem]{ulem}
\usepackage{sidecap}
\usepackage{hyperref}
\usepackage{cancel}
\usepackage{enumerate}
\hypersetup{pdfstartview=FitV,colorlinks=true,linkcolor=blue,citecolor=red,filecolor=black,urlcolor=blue}

\usepackage[T1]{fontenc}
\usepackage[utf8]{inputenc}
\usepackage{physics}

\def\stacksymbols #1#2#3#4{\def\theguybelow{#2}
    \def\vp{\lower#3pt}
    \def\sp{\baselineskip0pt\lineskip#4pt}
    \mathrel{\mathpalette\intermediary#1}}

\def\intermediary#1#2{\vp\vbox{\sp
     \everycr={}\tabskip0pt
     \halign{$\mathsurround0pt#1\hfil##\hfil$\crcr#2\crcr
              \theguybelow\crcr}}}

\newcommand{\lslashslash}{%

  \raisebox{0.8ex}{%
    \scalebox{.7}{%
      \rotatebox[origin=c]{18}{$-$}%
    }%
  }%
}
\newcommand{\lslash}{%
  {%
   \vphantom{d}%
   \ooalign{\kern-.1em\smash{\lslashslash}\hidewidth\cr${\rm l}$\cr}%
   \kern.05em
  }%
}

\newcommand*\diff{\mathop{}\!\mathrm{d}}

\newcommand{\beq}{\begin{equation}}
\newcommand{\eeq}{\end{equation}}
\newcommand{\bea}{\begin{eqnarray}}
\newcommand{\eea}{\end{eqnarray}}

\newcommand{\gsim}{\lower.7ex\hbox{$\;\stackrel{\textstyle>}{\sim}\;$}}
\newcommand{\lsim}{\lower.7ex\hbox{$\;\stackrel{\textstyle<}{\sim}\;$}}

\newcommand{\trh}{T_{\mathrm{RH}}}
\newcommand{\tmax}{T_{\mathrm{max}}}

\def\rhorh{\rho_{\rm RH}}
\def\arh{a_{\rm RH}}
\def\trh{T_{\rm RH}}
\def\ae{a_{\rm end}}
\def\rhoe{\rho_{\rm end}}

\def\be{\begin{equation}}
\def\ee{\end{equation}}
\def\bea{\begin{eqnarray}}
\def\eea{\end{eqnarray}}

\def\m{\mu}
\def\n{\nu}

\def\sp{\;\;\;,\;\;\;}

\def\lsim{\raise0.3ex\hbox{$\;<$\kern-0.75em\raise-1.1ex\hbox{$\sim\;$}}}
\def\gsim{\raise0.3ex\hbox{$\;>$\kern-0.75em\raise-1.1ex\hbox{$\sim\;$}}}

\def\inbar{\,\vrule height1.5ex width.4pt depth0pt}

\def\IC{\relax\hbox{$\inbar\kern-.3em{\rm C}$}}
\def\IQ{\relax\hbox{$\inbar\kern-.3em{\rm Q}$}}
\def\IR{\relax{\rm I\kern-.18em R}}
 \font\cmss=cmss10 \font\cmsss=cmss10 at 7pt
\def\IZ{\relax\ifmmode\mathchoice
 {\hbox{\cmss Z\kern-.4em Z}}{\hbox{\cmss Z\kern-.4em Z}}
 {\lower.9pt\hbox{\cmsss Z\kern-.4em Z}}
 {\lower1.2pt\hbox{\cmsss Z\kern-.4em Z}}\else{\cmss Z\kern-.4em Z}\fi}

\def\tmax{T_{\rm max}}
\def\trh{T_{\rm RH}}

\def\comment#1{}
\def\to{\rightarrow}

\def\u1x{U(1)_X}
\newcommand{\nc}{\newcommand}
\nc{\LL}{L}
\nc{\vv}{\tilde{v}}
\nc{\ccdot}{\!\cdot\!}
\nc{\gsm}{G_{SM}}
\nc{\vfive}{\mathbf{5}\oplus\mathbf{\overline{5}}}
\nc{\vten}{\mathbf{10}\oplus\mathbf{\overline{10}}}
\nc{\zhol}{Z^{\rm hol}}
\nc{\xfb}{\,{\rm fb}}

\begin{document}

%
%

\preprint{UMN--TH--4116/22}
\preprint{FTPI--MINN--22/07}
\preprint{CERN-TH-2022-025}

\vspace*{1mm}

\title{Gravitational Portals with Non-Minimal Couplings}

\author{Simon Cléry$^{a}$}
\email{simon.clery@ijclab.in2p3.fr}
\author{Yann Mambrini$^{a,b}$}
\email{yann.mambrini@ijclab.in2p3.fr}
\author{Keith A. Olive$^{c}$}
\email{olive@physics.umn.edu}
\author{Andrey Shkerin$^{c}$}
\email{ashkerin@umn.edu}
\author{Sarunas Verner$^{c}$}
\email{nedzi002@umn.edu}

\vspace{0.1cm}

\affiliation{
${}^a$ Universit\'e Paris-Saclay, CNRS/IN2P3, IJCLab, 91405 Orsay, France
 }
 \affiliation{
${}^b$ CERN, Theoretical Physics Department, Geneva, Switzerland
 }
 \affiliation{
${}^c$ 
 William I.~Fine Theoretical Physics Institute, 
       School of Physics and Astronomy,
            University of Minnesota, Minneapolis, MN 55455, USA
}

\begin{abstract} 

We consider the effects of non-minimal couplings 
to curvature of the form $\xi_S S^2 R$, for three types of scalars: the Higgs boson, the inflaton, and a scalar dark matter candidate. 
We compute the abundance of dark matter produced by these non-minimal couplings to gravity and compare to similar results with minimal couplings. We also compute the contribution to
the radiation bath during reheating. The main
effect is a potential augmentation of the maximum
temperature during reheating. A model independent limit of $\mathcal{O}(10^{12})$ GeV is obtained. For 
couplings $\xi_S \gtrsim \mathcal{O}(1)$, these dominate over minimal gravitational interactions.

\end{abstract}

\maketitle



\section{Introduction}

Promoting a field theory Lagrangian from a Lorentz-invariant one to a generally-covariant one necessarily leads to an interaction between the fields of the theory and the gravitational field. 
In the case of a scalar field, $S$, the natural generalization of this minimal interaction scenario is to introduce a non-minimal coupling term of the form 
\begin{equation}
\label{eq:gen_nonmin_int}
    \propto \xi_S S^2R \;.
\end{equation}
Here $R$ is the Ricci scalar and $\xi_S$ is a non-minimal coupling constant. 
This non-minimal coupling to gravity proved to be useful in many applications to cosmology. Examples include Higgs inflation \cite{Bezrukov:2007ep,Lebedev:2021xey}, where $S$ is associated with the Higgs field degree of freedom $h$ --- the only scalar degree of freedom in the Standard Model, preheating \cite{Ema:2016dny}, where $S$ is associated with the inflaton field $\phi$, and non-perturbative production of dark matter \cite{nonminprod}, where $S$ represents the scalar dark matter particle $X$. 

In the general case, when the fields $\phi$, $h$, and $X$ are all different, the question arises as to what extent they must interact with each other in order to successfully reheat the Universe and generate the right amount of dark matter. 
Recent studies have shown that interactions via gravity alone, to which the fields are coupled minimally, is enough for these purposes. Indeed, the perturbative gravitational production of dark matter through graviton exchange can play a dominant role during reheating with processes involving the inflaton \cite{MO,CMOV,Barman:2021ugy} as well as thermal bath particles \cite{CMOV,Haque:2021mab}. 
Further, the minimal gravitational coupling can lead to the completion of the reheating process for certain types of the inflationary potential, $V(\phi) \sim \phi^k$ with $k > 2$ \cite{CMOV,Haque:2022kez}. Thus, gravity is strong enough to mediate perturbative channels of reheating and dark matter production.

The purpose of this work is to study how the inclusion of the non-minimal coupling terms of the form (\ref{eq:gen_nonmin_int}) affect the gravitational production of dark matter and radiation during reheating. 
Note that the presence of these terms is unavoidable: if there were no such couplings at tree level, they would still be generated by quantum corrections \cite{Callan:1970ze}. 
We study particle production in the processes $hh\to XX$, $\phi\phi\to hh$, and $\phi\phi\to XX$ which are induced by the non-minimal couplings.
Here $\phi$ represents the inflaton background oscillating around its minimum after the end of inflation~\cite{gravprod}. 
Since the scalar fields couple directly to the curvature scalar $R$, the oscillating background causes the effective masses of the fields to change non-adiabatically and leads to particle production. 
This regime of particle creation has been considered in several different contexts, including gravitational production of scalar~\cite{gravscalar,ema}, fermion~\cite{gravferm}, and vector dark matter~\cite{gravvector}.

Our main interest is to compare the (dark) matter production channels induced by the non-minimal couplings with the production via the s-channel graviton exchange that sets minimal possible production rates. We will see for which values of the couplings the rates are enhanced, and what are the consequences on the dark matter density or the temperature attained during reheating.
Throughout the work we adopt the Starobinsky inflationary potential~\cite{staro}, although our results are largely independent of the particular form of the potential. 
As for the potentials for the fields $h$ and $X$, we take them to be renormalizable polynomials. We also assume no direct interaction between $\phi$, $h$, and $X$.

Working in the perturbative regime implies that the non-minimal couplings must satisfy $|\xi_S| \ll M_P^2/\langle S \rangle^2$, where $\langle S \rangle$ is the vacuum expectation value of $S=\phi,h,X$. The value of $\xi_h$ is constrained from collider experiments as $|\xi_h|\lesssim 10^{15}$~\cite{higgscons1}.\footnote{Note that in the case of Higgs inflation, $\xi_h$ is fixed from CMB measurements~\cite{Bezrukov:2007ep}.} Furthermore, the lower bound on $\xi_h$ comes from the fact that the Standard Model electroweak vacuum may not be absolutely stable~\cite{HiggsStab}. To prevent the vacuum decay due to quantum fluctuations during inflation~\cite{fluc}, the effective mass of the Higgs field induced by the non-minimal coupling must be large enough; this gives $\xi_h\gtrsim 10^{-1}$~\cite{higgstree,higgsloop} (see also \cite{Markkanen:2018pdo}).\footnote{This estimate assumes no new physics interfering the RG running of the Higgs self-coupling constant until inflationary energy scales.} 

The paper is organized as follows: The framework for our computation is presented
in Section~\ref{Sec:framework}. We discuss non-minimal gravitational couplings of the inflaton, the Higgs boson, and a dark matter scalar in detail. We calculate the dark matter production rates either from scattering in the thermal bath or from oscillations in the inflaton condensate. We compare similar processes obtained from the minimal gravitational particle production. We choose the Starobinsky model of inflation and discuss the reheating epoch when the inflaton begins oscillating. In Section~\ref{Sec:dm} we discuss the resulting abundance of dark matter produced from the thermal bath and directly from scattering of the inflaton condensate. We also compute the effects of the non-minimal couplings on the maximum temperature attained during reheating. We then compare different processes in Section~\ref{Sec:results}, before summarizing our results in Section~\ref{Sec:conclusion}.

\section{The framework}
\label{Sec:framework}

\subsection{Scalar-gravity Lagrangian}
\label{Ssec:lagr}

The theory we consider comprises 3 scalar fields non-minimally coupled to gravity: the inflaton $\phi$, the Higgs field\footnote{We consider the Higgs boson as a surrogate for any additional scalars with Standard Model couplings.} $H$, for which we adopt the Unitary gauge, $H=(0,h)^T/\sqrt{2}$, and the dark matter candidate $X$. 
The relevant part of the action takes the form\footnote{The metric signature is chosen as $(+,-,-,-)$.}
\begin{equation}
    \mathcal{S} \; = \; \int d^4 x \sqrt{-\tilde{g}} \left[-\frac{M_P^2}{2} \Omega^2 \tilde{R} +\mathcal{L}_{\phi}  + \mathcal{L}_{h} + \mathcal{L}_{X} \right]
    \label{eq:jordan}
\end{equation}
with the conformal factor $\Omega^2$ given by
\begin{equation}
    \label{eq:conformalfact}
    \Omega^2 \; \equiv \; 1 + \frac{\xi_{\phi} \phi^2}{M_P^2} + \frac{\xi_{h} h^2}{M_P^2} + \frac{\xi_{X} X^2}{M_P^2} \, .
\end{equation}
Here $M_P=2.4\times 10^{18}\,\rm{GeV}$ is the reduced Planck mass, and the tilde used in Eq.~(\ref{eq:jordan}) indicates that the theory is considered in the Jordan frame. 
For the scalar field Lagrangians we have
\begin{equation}
\label{eq:L_S}
    \mathcal{L}_{S} \; = \frac{1}{2} \tilde{g}^{\mu\nu} \partial_{\mu} S \partial_{\nu} S - V_{S} \, , ~~~ S=\phi,h,X \, .
\end{equation}

Next, we specify the scalar field potentials. For a model of inflation, we choose the well-motivated Starobinsky model for which~\cite{staro} 
\begin{equation}
    \label{eq:inflpot}
    V_{\phi} \; = \; \frac{3}{4} m_{\phi}^2 M_P^2 \left(1 - e^{-\sqrt{\frac{2}{3}} \frac{\phi}{M_P}} \right)^2 \, .
\end{equation}
In what follows, we work in the perturbative regime with $\phi\ll M_P$, hence the potential is approximated as
\begin{equation}
\label{eq:inflpot_approx}
    V_{\phi} \; \simeq \; \frac{1}{2} m_\phi ^2 \phi^2 \, .
\end{equation}
The inflaton mass, $m_\phi$, is fixed by the amplitude of scalar perturbations inferred from CMB measurements~\cite{Planck}; for the potential (\ref{eq:inflpot}) this gives $m_\phi = 3 \times 10^{13}$ GeV~\cite{building}. 

The potential for the Higgs field is taken as follows
\begin{equation}
\label{eq:HiggsPot}
    V_h \; = \; \frac{1}{2} m_h^2 h^2+\frac{1}{4} \lambda_h h^4   \, .
\end{equation}
Here $m_h$ and $\lambda_h$ are the Higgs mass and quartic coupling, correspondingly. Note that both parameters undergo the renormalization group (RG) running. In what follows we take a weak scale mass, which is a good approximation at the time of reheating and our results are insensitive to $\lambda_h$. Finally, the dark matter potential is simply given by
\begin{equation}
\label{eq:DMpot}
    V_X \; = \; \frac{1}{2} m_X^2 X^2 \, .
\end{equation}

To study the reheating in the theory (\ref{eq:jordan}), it is convenient to remove the non-minimal couplings by performing the redefinition of the metric field. 
Leaving the details to Appendix~\ref{sec:appendixA}, we write the action (\ref{eq:jordan}) in the Einstein frame,
\begin{equation}
    \label{eq:einstein}
\begin{split}
    \mathcal{S} \; = \; \int d^4 x \sqrt{-g} & \left[-\frac{M_P^2}{2}R + \frac{1}{2} K^{ij} g^{\mu\nu} \partial_{\mu}S_i \partial_{\nu}S_j \right. \\
    & \qquad\qquad\qquad \left. - \frac{V_\phi+V_h+V_X}{\Omega^4} \right] \, .
\end{split}
\end{equation}
Here the indices $i,j$ enumerate the fields $\phi, h, X$, and the kinetic function is given by
\begin{equation}
    \label{eq:kinfunc}
    K^{i j} \; = \; 6 \frac{\partial \log \Omega}{\partial S_i} \frac{\partial \log \Omega}{\partial S_j} + \frac{\delta^{ij}}{\Omega^2} \, .
\end{equation}
Note that the scalar field kinetic term is not canonical. 
In general, it is impossible to make a field redefinition that would bring it to the canonical form, unless all three non-minimal couplings vanish.\footnote{ Such a redefinition exists if the three-dimensional manifold spanned by the fields $\phi$, $h$ and $X$ is flat. One can show that it is not the case if at least one of the couplings is non-zero.} 
For the theory~(\ref{eq:einstein}) to be well-defined, the kinetic function (\ref{eq:kinfunc}) must be positive-definite. Computing the eigenvalues, one arrives at the condition
\begin{equation}
    \Omega^2 > 0 \, ,
\end{equation}
which is satisfied automatically for positive values of the couplings. Note that the negative couplings are also allowed for certain scalar field magnitudes.

In what follows, we will be interested in the small-field limit
\begin{equation}
    \label{eq:smallfield}
    \frac{|\xi_{\phi}| \phi^2}{M_P^2} \;, ~~ \frac{|\xi_{h}| h^2}{M_P^2} \;, ~~ \frac{|\xi_{X}| X^2}{M_P^2} \ll 1 \, .
\end{equation}
We can expand the kinetic and potential terms in the action (\ref{eq:einstein}) in powers of $M_P^{-2}$. We obtain a canonical kinetic term for the scalar fields and deduce the leading-order interactions induced by the non-minimal couplings. 
The latter can be brought to the form 
\begin{equation}
    \label{lag4point}
    \mathcal{L}_{\rm{non-min.}} \; = \; -\sigma_{hX}^{\xi} h^2 X^2 - \sigma_{\phi X}^{\xi} \phi^2 X^2 - \sigma_{\phi h}^{\xi} \phi^2 h^2 \, , 
\end{equation}
where the $\sigma_{ij}^{\xi}$ are functions of the couplings $\xi_i$, $\xi_j$, the masses $m_i$, $m_j$, and the Mandelstam variables; see Appendix~\ref{sec:appendixA} for details. 

The small-field approximation~(\ref{eq:smallfield}) implies the bound $\sqrt{|\xi_{S}|} \lesssim M_P/\langle S\rangle$ with $S = \phi, h, X$. 
Since the inflaton value at the end of inflation is $\phi_{\rm end}\sim M_P$ and afterwards $\langle \phi^2 \rangle  \sim a^{-3} $, where $a$ is the cosmological scale factor, then $|\xi_\phi|\lesssim (a/a_{\rm end})^3$. 
In particular, at the onset of inflaton oscillations 
\begin{equation}
\label{eq:bound_on_xi_phi}
   | \xi_\phi | \lesssim 1 \, .
\end{equation}
Note that since our calculations involve the effective couplings $\sigma_{\phi X}^{\xi}$ ($\sigma_{\phi h}^{\xi}$), which depend both on $\xi_\phi$ and $\xi_X$ ($\xi_h$), the relatively small value of $| \xi_{\phi} |$ can, in principle, be compensated by a large value of the other couplings.

\begin{figure}[ht!] 
    \includegraphics[width = 0.6\columnwidth]{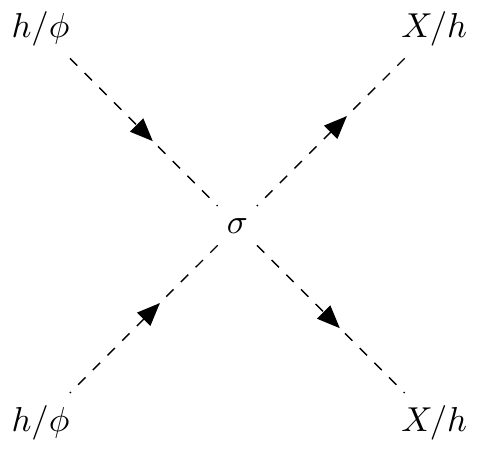}
    \caption{\em \small Feynman diagram for the 4-point interactions between the inflaton $\phi$, the dark matter scalar candidate $X$, and the Higgs boson $h$, given by the Lagrangian~(\ref{lag4point}).}
    \label{fig:feyn1}
\end{figure}

In Fig.~\ref{fig:feyn1}, we show the scattering processes obtained from the Lagrangian~(\ref{lag4point}).
These contribute to reheating (when $h$ is in the final state) and dark matter production (when $X$ is in the final state). 

Finally, in evaluating the cosmological parameters, it is important to stay within the validity of the low-energy theory. The cutoff of the theory can be estimated as (see, e.g., \cite{higgscons2})
\begin{equation}
    \Lambda\sim\frac{M_P}{\text{max}_i \:|\xi_i| } \, .
\end{equation}
In particular, the temperature of reheating must not exceed $\Lambda$.

\subsection{Graviton exchange}
\label{Ssec:gravExch}

\begin{figure}[ht!] 
    \includegraphics[width = 1\columnwidth]{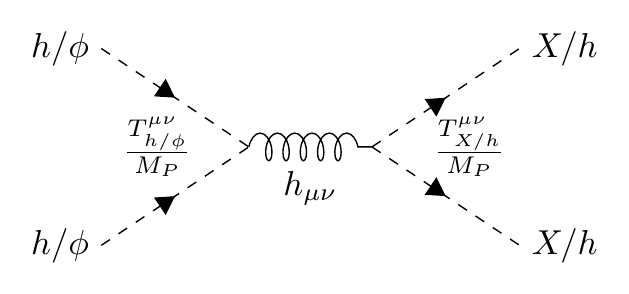}
    \caption{\em \small Feynman diagram for the (dark) matter production through the gravitational scattering of the inflaton or the Higgs boson from the thermal bath.
    }
    \label{fig:feyn2}
\end{figure}

Let us first consider the case of vanishing $\xi_{\phi, \, h, \, X}$, i.e., the case of the minimal coupling of the scalar fields to gravity~\cite{MO, CMOV,ema,Garny:2015sjg,Tang:2017hvq,Chianese:2020yjo,Redi:2020ffc}. 
It was argued in~\cite{MO,CMOV} that the interaction between the dark and visible sectors induced by gravity leads to unavoidable contributions to reheating and dark matter production, in the thermal bath or via the scattering of the inflaton condensate, through the graviton exchange processes shown in Fig.~\ref{fig:feyn2}.  
It is therefore important to compare the minimal gravitational particle
production to similar processes obtained from the
Lagrangian in Eq.~(\ref{lag4point}) with non-minimal couplings.

To study the universal gravitational interactions in minimally coupled gravity, we expand the space-time metric around flat space using $g_{\mu \nu} \simeq \eta_{\mu \nu} + 2 h_{\mu \nu}/M_P$, where $h_{\mu \nu}$ is the canonically-normalized perturbation. The gravitational interactions are characterized by the following Lagrangian,
\beq
    {\cal L}_{\rm min.}= -\frac{1}{M_P}h_{\mu \nu}
    \left(T^{\mu \nu}_{h}+T^{\mu \nu}_\phi + T^{\mu \nu}_{X} \right) \, ,
    \label{eq:lagrgrav}
\eeq
where the stress-energy tensor is given by
\begin{equation}
    T^{\mu \nu}_{S} \; = \; 
\partial^\mu S \partial^\nu S-
g^{\mu \nu}
\left[
\frac{1}{2}\partial^\alpha S \partial_\alpha S-V_S \right] \, .
\label{eq:tensors}
\end{equation}
Note that in this work, we consider only the Higgs field in the visible sector.
Generalization to the complete spectrum of the Standard Model is
straightforward, and we leave it for future work.

For models with minimally coupled gravity, the processes $\phi/h(p_1)+\phi/h(p_2) \rightarrow {h}/X(p_3)+{h}/X(p_4)$ can be parametrized by
\begin{equation}
\label{eq:partialamp}
\mathcal{M}^{00} \propto M_{\mu \nu}^0 \Pi^{\mu \nu \rho \sigma} M_{\rho \sigma}^0 \;, 
\end{equation}
where the graviton propagator for the canonically-normalized field $h_{\mu\nu}$ with exchange momentum $k = p_1 + p_2$ is given by
\begin{equation}
 \Pi^{\m\n\rho\sigma}(k) = \frac{\eta^{\m \rho}\eta^{\n\sigma} + 
\eta^{\m \sigma}\eta^{\n \rho} - \eta^{\m\n}\eta^{\rho\sigma} }{2k^2} \, ,
\end{equation} 
and the partial amplitude, $M_{\mu \nu}^0$, is given by
\begin{flalign}
\label{eq:partialamp2}
M_{\mu \nu}^{0} = \frac{1}{2}\left[p_{1\mu} p_{2\nu} + p_{1\nu} p_{2\mu} - \eta_{\mu \nu}p_1\cdot p_2 - \eta_{\mu \nu} V_S''\right] \, ,
\end{flalign}
with analogous expression for the final state in terms of outgoing momenta $p_{3, 4}$ and the final state potential. 
In Fig.~\ref{fig:feyn2} we show the s-channel graviton exchange scattering obtained from the Lagrangian~(\ref{eq:lagrgrav}) for the production of dark matter from either the Higgs field or the inflaton condensate as well as the reheating process (the production of Higgs bosons from the inflaton condensate).  

\subsection{Production rates}

In this work, we consider three processes: 

\begin{enumerate}[A.]
\item{The production of dark matter from the 
scattering of thermal Higgs bosons 
(assuming reheating is produced by inflaton decay). 
In this case, the dark matter  
is populated via a freeze-in mechanism throughout the reheating period.} 

\item{The production of dark matter from  
direct excitations
of the inflaton condensate. 
This process, which can be viewed as gravitational 
inflaton scattering,
is independent of the presence of a thermal bath.}

\item{The creation of a radiative bath at the start of reheating arising from the Higgs boson production through  gravitational inflaton scattering. Since such a process is unavoidable in 
minimally coupled gravity, 
it  is interesting to know when such a process becomes
dominant in models with non-minimal couplings $\xi_i$.}
\end{enumerate}

The thermal dark matter production rate $R(T)$ for the process $h h \rightarrow X X$ can be calculated from\footnote{We include the symmetry factors associated with identical initial and final states in the definition of $|\overline{{\cal{M}}}|^2$, and a factor of 2 is explicitly included in the definition of the rate to account for the production of 2 identical particles.}  \cite{gravitino}
\beq
    \label{eq:rategen}
    R(T) = \frac{2 \times N_h}{1024 \pi^6}\int f_1 f_2 E_1 \diff E_1 E_2 \diff E_2 \diff \cos \theta_{12}\int |\overline{{\cal M}}|^2 \diff \Omega_{13} \, ,
\eeq
where $E_i$ is the energy of particle $i=1,2$, $\theta_{13}$ and $\theta_{12}$ are the angles formed by momenta ${\bf{p}}_{1,3}$ and ${\bf{p}}_{1,2}$, respectively. $N_h = 4$ is the number of internal degrees of freedom for 1 complex Higgs doublet, $|\overline{{\cal M}}|^2$ is the matrix amplitude squared with all symmetry factors included. This accounts for the explicit factor of 2 in the numerator of Eq.~(\ref{eq:rategen}). The thermal distribution function of the incoming Higgs particles is given by the Bose-Einstein distribution
\beq
\label{eq:boseeinstein}
f_{i}= \frac{1}{e^{E_i/T} - 1} \, .
\eeq
The rate for minimal gravitational interactions from Eq.~(\ref{eq:lagrgrav})
was derived in \cite{CMOV,Bernal:2018qlk}.  The rate we use here differs in two respects. As noted earlier, we only include Higgs scalars in the initial state whereas in \cite{CMOV,Bernal:2018qlk}, all Standard Model particle initial states were included. Secondly, we keep terms depending on the dark matter mass which had not previously been taken into account. This allows us to consider dark matter masses approaching the inflaton mass and/or the reheating temperature.

\begin{figure}[ht!] 
    \includegraphics[width = 1.0\columnwidth]{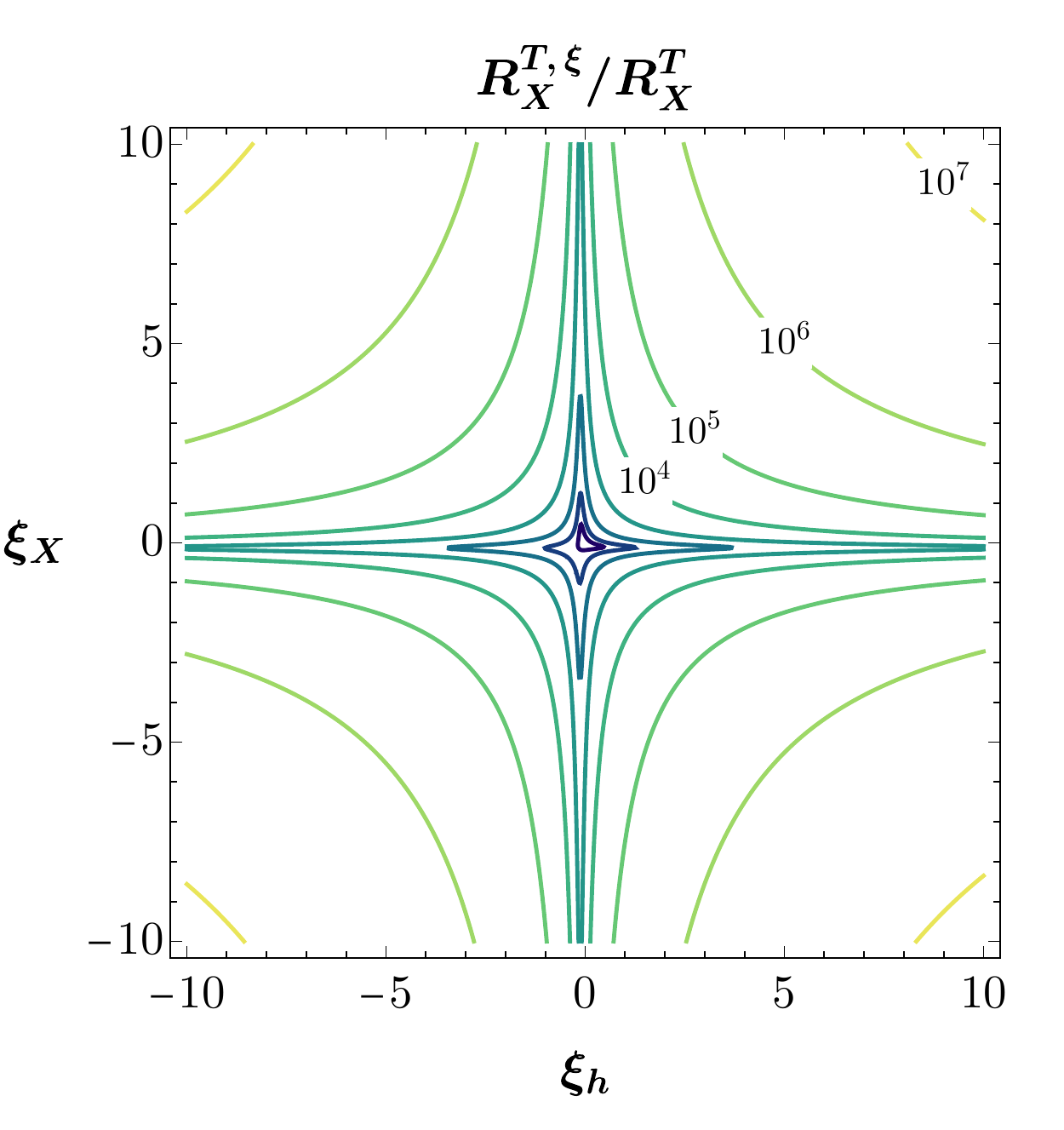}
\caption{\em \small Contours of the ratio of
the dark matter production rates from the thermal bath based on non-minimal gravitational interactions to those based on minimal interactions. The ratio is displayed in the $(\xi_h,\xi_X)$ plane.  Note that as discussed in the Introduction, negative values of $\xi_h$ may require new physics (such as supersymmetry) to stabilize the Higgs vacuum. }
    \label{plotsxi1}
\end{figure}

For minimal (non-minimal) gravitational interactions, we find that the thermal dark matter production rate  can be expressed as
\begin{flalign}
    \label{eq:ratefull1}
    R_X^{T, \, (\xi)} (T) & = & \beta_1^{(\xi)} \frac{T^8}{M_P^4} + \beta_2^{(\xi)} \frac{m_X^2 T^6}{M_P^4} + \beta_3^{(\xi)} \frac{m_X^4 T^4}{M_P^4} \, ,
\end{flalign}
where the coefficients $\beta_{1, \,  2, \, 3}^{(\xi)}$ are given in Appendix~\ref{sec:appendixB} by Eqs.~(\ref{eq:b1}-\ref{eq:b3})
(Eqs.~(\ref{eq:b1xi}-\ref{eq:b3xi})).
The ratio of the non-minimal to minimal rate is shown in Fig.~\ref{plotsxi1}. However, we note that when $\xi_i \sim \mathcal{O}(1)$ both rates are comparable and interference effects become significant. The full coefficients $\beta_{1,\, 2,\, 3}$ including interference are given by Eqs.~(\ref{eq:b11}-\ref{eq:b33}) from Appendix~\ref{sec:appendixB}. We leave the comparison of the effects on dark matter production from the two rates for the next section. 

The rate for dark matter produced from inflaton oscillations of the inflaton condensate
for a potential of the form $V = \lambda \phi^k$ were considered in detail in \cite{GKMO2,CMOV}. The time-dependent inflaton can be written as $\phi(t) = \phi_0(t)  \mathcal{P}(t)$, where $\phi_0(t)$ is the time-dependent amplitude that includes the effects of redshift and $\mathcal{P}(t)$ describes the periodicity of the oscillation.
The dark matter production rate is calculated by writing the potential in terms of the Fourier modes of the oscillations \cite{Ichikawa:2008ne,Kainulainen:2016vzv,GKMO2,CMOV}
\beq
V(\phi)=V(\phi_0)\sum_{n=-\infty}^{\infty} {\cal P}_n^ke^{-in \omega t}
=\rho_\phi\sum_{n = -\infty}^{\infty} {\cal P}_n^ke^{-in \omega t} \, .
\eeq
For $k=2$ (the only case considered here),
the frequency of oscillation is simply,
$\omega = m_\phi$.

The rate generated by non-minimal couplings
can be readily calculated using the Lagrangian~(\ref{lag4point}), which leads to 
\begin{equation}    
    \label{eq:rateinfdm0}
    R_X^{\phi,  \, \xi} \; = \; \frac{2 \times \sigma_{\phi X}^{\xi~2}}{\pi} \frac{ \rho_{\phi}^2}{m_{\phi}^4} \Sigma_0^k \, ,
\end{equation}
where
\begin{equation}
   \Sigma_0^k = \sum_{n = 1}^{\infty}  |{\cal P}^k_n|^2
\sqrt{1-\frac{4m_X^2}{E_n^2}} \, ,
\label{Sigma0k}
\end{equation}
and $E_n = n \omega$ is the energy of the $n$-th inflaton oscillation mode.
For $k=2$, only the second Fourier mode in the sum contributes, with $\sum |\mathcal{P}^2_n|^2 = \frac{1}{16}$. Thus, the rate becomes
\begin{equation}    
    \label{eq:rateinfdm1}
    R_X^{\phi,  \, \xi} \; = \; \frac{2 \times \sigma_{\phi X}^{\xi~2}}{16\pi} \frac{ \rho_{\phi}^2}{m_{\phi}^4} \sqrt{1 - \frac{m_X^2}{m_{\phi}^2}}\, ,
\end{equation}
where $\rho_{\phi}$ is the energy density of the inflaton and the interaction term $\sigma_{\phi X}^{\xi}$ is given in Appendix~\ref{sec:appendixA} by Eq.~(\ref{appa:sigphiX}).

\begin{figure}[t!] 
    \includegraphics[width = 1.0\columnwidth]{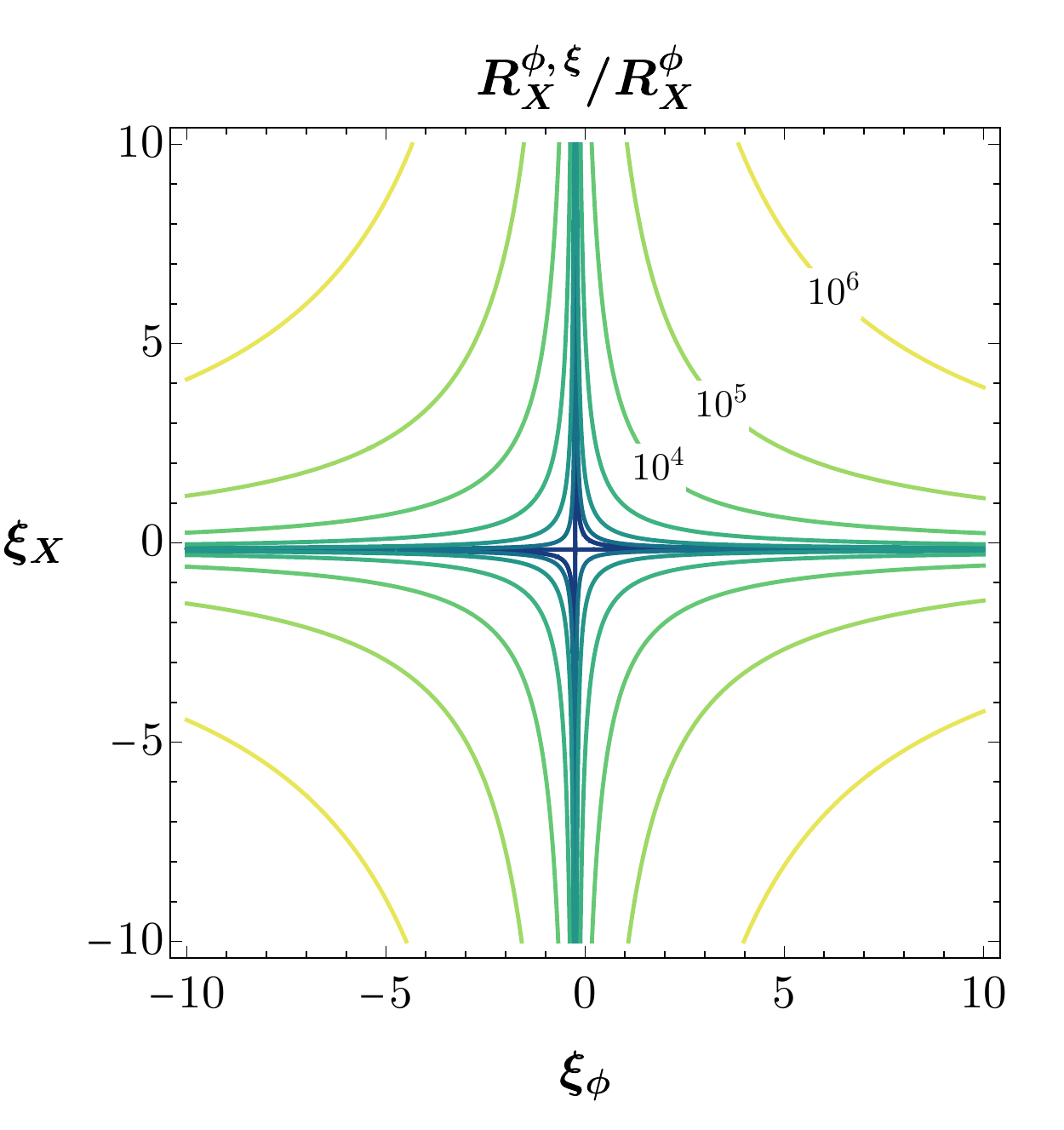}
    \caption{\em \small Contours of the ratio of
the dark matter production rates from oscillations in the inflaton condensate based on non-minimal gravitational interactions to those based on minimal interactions. The ratio is displayed in the $(\xi_\phi,\xi_X)$ plane.}
    \label{plotsxi2}
\end{figure}

It was shown in~\cite{MO} that the dark matter production rate through 
the exchange of a graviton, computed from the partial amplitude~(\ref{eq:partialamp}), is
\beq
\label{eq:ratescalar2}
R^{\phi}_X=\frac{ 2 \times \rho_\phi^2}{256 \pi M_P^4}
\left(1+\frac{m_X^2}{2m^2_\phi}\right)^2\sqrt{1-\frac{m_X^2}{m_\phi^2}} \, ,
\eeq
which can be written in the same form as (\ref{eq:rateinfdm1}) by defining an effective coupling $\sigma_{\phi X}$
\begin{equation}
    \sigma_{\phi X} \; = \; -\frac{m_{\phi}^2}{4M_P^2} \left(1 + \frac{m_X^2}{2m_{\phi}^2} \right) \, .
\end{equation}
A comparison of the non-minimal to minimal rates for the production of dark matter from inflaton scattering is shown in Fig.~\ref{plotsxi2}.

For the production of Higgs bosons through inflaton condensate scattering, we follow a similar procedure, and from the Lagrangian~(\ref{lag4point}) we find 
\begin{equation}
    \label{eq:ratehiggs}
    R_h^{\phi,  \, \xi} \; \simeq \; N_h \frac{2 \times \sigma_{\phi h}^{\xi~2}}{16 \pi} \frac{ \rho_{\phi}^2}{m_{\phi}^4} \, ,
\end{equation}
where we assumed that $m_h \ll m_{\phi}$, $N_h = 4$ is the number of internal degrees of freedom for 1 complex Higgs doublet, and $\sigma_{\phi h}^{\xi}$ is given in Appendix~\ref{sec:appendixA} by Eq.~(\ref{appa:sigphih}).

On the other hand, it was argued in~\cite{CMOV} that the  scattering 
 $\phi \phi \rightarrow h h$ through the graviton exchange can 
 also be parameterized by an effective coupling 
\begin{equation}
   \mathcal{L}_h = - \sigma_{\phi h} \phi^2 h^2 \, , 
\end{equation}
with
\begin{equation}
    \sigma_{\phi h} \; = \; -\frac{m_{\phi}^2}{4 M_P^2} \, ,
    \label{sph}
\end{equation}
and the rate $R_h^{\phi}$ is given by the analogous expression to~(\ref{eq:ratehiggs}) with $\sigma_{\phi h}^{\xi}$ replaced by $\sigma_{\phi h}$.

The full four-point coupling of course is given by
the sum $\sigma_{\phi h/X}^{\xi} + \sigma_{\phi h/X}$.
However, except for values where the two are similar, which occurs when $12 \xi^2 + 5 \xi \simeq \frac12$ (assuming $m_X \ll m_\phi$ and taking the $\xi_i$ to be equal to $\xi$), either the minimal or the non-minimal contribution dominates. Thus, for the most part, we will consider separately the minimal and non-minimal contributions. Note that for two values of $\xi$ ($\xi \sim -1/2$ and 1/12) destructive interference could occur causing the entire rate to vanish (at the tree level). 

\section{Particle Production with a Non-Minimal Coupling}
\label{Sec:dm}
Given the rates $R_i^j$ calculated in the previous section, we compute the 
evolution for the gravitational (minimal and non-minimal) 
contribution to the reheating processes 
and the dark matter density for the three reactions outlined above.

\subsection{$h~h \rightarrow X~X$}
The gravitational scattering of thermal Higgs bosons leads to the production of massive scalar dark matter particles $X$. The dark matter number density $n_X$ can be calculated from the classical Boltzmann equation
\beq
\frac{dn_X}{dt} + 3Hn_X = R^T_X \,,
\label{eq:boltzmann1}
\eeq
where $H=\frac{\dot a}{a}$ is the Hubble parameter and the right-hand side of the equation represents the dark matter production rate. 
It is more practical to rewrite the above equation in terms of the scale factor $a$ rather than the parameters $t$ or $T$. 

 We proceed by introducing the comoving number density $Y_X= n a^3$ and rewriting the Boltzmann equation as
\beq
\frac{dY_X}{da}=\frac{a^2R^T_X(a)}{H(a)} \, .
\label{eq:boltzmann3}
\eeq
Since the production rate (\ref{eq:ratefull1}) is a function of the temperature of the thermal bath, it is necessary to determine the relation between $T$ and $a$ 
in order to solve the Boltzmann equation as a function of the scale factor $a$.
For the Starobinsky potential in Eq.~(\ref{eq:inflpot}), at the end of inflation, the inflaton starts oscillating about a quadratic minimum, and we find the following energy conservation equations\footnote{For the inflaton scattering with $V(\phi) \sim \phi^k$, where $k >2$, see \cite{Bernal:2019mhf,GKMO1,GKMO2,GKMOV,Haque:2021mab,Haque:2022kez,Ahmed:2021fvt}.}
\bea
&&
\frac{d \rho_\phi}{dt} + 3H\rho_\phi = -\Gamma_\phi \rho_\phi \, ,
\label{eq:diffrhophi}
\\
&&
\frac{d \rho_R}{dt} + 4H\rho_R = \Gamma_\phi \rho_\phi \, ,
\label{eq:diffrhor}
\eea
where $\rho_{\phi}$ and $\rho_R$ are the energy density of the
inflaton and radiation, respectively, $\Gamma_{\phi}$ is the inflaton decay rate, and for a quadratic minimum, we are able to set the equation of state parameter $w_{\phi} = \frac{P_{\phi}}{\rho_{\phi}}\simeq 0$. 
We will assume that reheating occurs 
due to an effective inflaton coupling to the Standard Model fermions, given by the interaction Lagrangian
\begin{center}
\begin{equation}
    \mathcal{L}_{\phi-SM}^{y} \; = \; - y \phi \bar{f} f \, ,
\end{equation}
\end{center}
where $y$ is a Yukawa-like coupling, $f$ is a Standard Model fermion, and the inflaton decay rate is
\begin{equation}
    \Gamma_{\phi} \; = \; \frac{y^2}{8 \pi} m_{\phi} \, .
\end{equation}

If we solve the Friedmann equations~(\ref{eq:diffrhophi},~\ref{eq:diffrhor}), we find~\cite{GKMO1,GKMO2,CMOV}
\beq
\rho_\phi(a) = \rho_{\rm end} \left(\frac{a_{\rm end}}{a} \right)^3
\label{rhophia}
\eeq
and
\beq
\rho_R(a)=\rhorh\left(\frac{\arh}{a}\right)^{\frac{3}{2}}
\frac{1-\left(\frac{\ae}{a}\right)^{\frac{5}{2}}}
{1-\left(\frac{\ae}{\arh}\right)^{\frac{5}{2}}} \, ,
\label{Eq:rhoR}
\eeq
where $a_{\rm{end}}$ is the scale factor at the end of inflation, $\rhoe \equiv \rho_{\phi}(a_{\rm{end}})$ is the inflaton energy density at the end of inflation when there is no radiation present, $a_{\rm{RH}}$ is the scale factor at reheating, and $\rhorh \equiv \rho_R(a_{\rm{RH}}) = \rho_{\phi}(a_{\rm{RH}})$ is the energy density at reheating. We note that these equations are strictly valid for $a_{\rm end} \ll a \ll a_{\rm{RH}}$  and the end of inflation occurs when $\ddot{a} = 0$ which corresponds to $\rhoe = \frac{3}{2} V(\phi_{\rm{end}})$. For the Starobinsky potential, $\rhoe \simeq 0.175 m_\phi^2 M_P^2$~\cite{egno5}. 

The radiation energy density can be parameterized as
\beq
\rho_R=\frac{g_T\pi^2}{30}T^4\equiv \alpha T^4 \, ,
\eeq
where $g_T$ is the number of relativistic degrees of freedom at the temperature $T$. The maximum temperature is attained when the radiation energy density reaches its peak at $\rho_R(a_{\rm{max}}) = \alpha T_{\rm{max}}^4$. It was shown in~\cite{GKMO1} that the ratio of  $a_{\rm max}$ to $a_{\rm end}$ 
is given by
\beq
    \label{Eq:amax}
    \frac{a_{\rm max}}{\ae}=\left(\frac{8}{3}\right)^{\frac{2}{5}} \simeq 1.48 \, .
\eeq

Using Eq.~(\ref{Eq:rhoR}) we can then express the production rate
from gravitational scattering of thermal particles~(\ref{eq:ratefull1})
as a function of the scale factor $a$
\beq
R_{X}^{T, \,( \xi)}(a) \simeq \beta_1^{(\xi)} \frac{\rhorh^2}{\alpha^2M_P^4} \left(\frac{\arh}{a}\right)^{3}
\left[\frac{1 - \left(\frac{a_{\rm end}}{a} \right)^\frac{5}{2}}
{1 - \left(\frac{a_{\rm end}}{\arh} \right)^\frac{5}{2}}
\right]^2 \, ,
\label{rateXT}
\eeq
where we assumed that $m_{X} \ll m_{\phi},\,T$, and thus neglected the terms $\beta_{2, \, 3}^{(\xi)}$. If we use $H \simeq \frac{\sqrt{\rho_\phi(a)}}{\sqrt{3}M_P}$,
which is valid for $a \ll a_{\rm RH}$, we can rewrite Eq.~(\ref{eq:boltzmann3}) as
\beq
\frac{dY_X^{\xi}}{da}=\frac{\sqrt{3}M_P}{\sqrt{\rhorh}}a^2\left(\frac{a}{a_{\rm RH}}\right)^{\frac{3}{2}}R_X^{T, \, (\xi)}(a) \, .
\label{Eq:boltzmann3}
\eeq
We find that the solution to this equation is
\begin{flalign}
\label{Eq:nxthermal}
&~~n^{T, \, \xi}_X(a_{\rm RH})  = 
\frac{2 \beta_1^{\xi}}{\sqrt{3} \alpha^2M_P^3}
\frac{\rhorh^{3/2}}{(1-(\ae/\arh)^{\frac{5}{2}})^2} \times 
\nonumber \\
& \left(1 + 3 \left(\frac{a_{\rm end}}{a_{\rm RH}} \right)^\frac{5}{2} -  \frac{25}{7}\left(\frac{a_{\rm end}}{a_{\rm RH}} \right)^\frac{3}{2} - \frac{3}{7}\left(\frac{a_{\rm end}}{a_{\rm RH}} \right)^5\right) \, ,
\end{flalign}
where we integrated Eq.~(\ref{Eq:boltzmann3}) in the interval $a_{\rm end} < a < a_{\rm RH}$.

The relic abundance is given by~\cite{book}
\beq
\Omega_Xh^2 = 1.6\times 10^8\frac{g_0}{g_{\rm RH}}\frac{n(\trh)}{\trh^3}\frac{m_X}{1~{\rm GeV}} \, ,
\eeq
and if we combine it with Eq.~(\ref{Eq:nxthermal}), we obtain
\bea
    \Omega_X^{T, \, (\xi)} h^2 \; = \;  \frac{2}{3} \Omega_k^{(\xi)}  &&\left[1 + 3 \left( \frac{\rho_{\rm{RH}}}{\rho_{\rm{end}}}\right)^{\frac{5}{6}} 
    -\frac{25}{7} \left( \frac{\rho_{\rm{RH}}}{\rho_{\rm{end}}}\right)^{\frac{1}{2}} 
    \right.
    \nonumber
    \\ 
    &&- \left.
    \frac{3}{7}  \left( \frac{\rho_{\rm{RH}}}{\rho_{\rm{end}}}\right)^{\frac{5}{3}} \right] \, ,
    \label{oh2Txi}
\eea
with
\beq
\Omega_k^{(\xi)} = 1.6 \times 10^8 \frac{g_0}{g_{\rm RH}}
    \frac{\beta_1^{(\xi)}\sqrt{3}}{\sqrt{\alpha} }
    \frac{m_X}{1~\rm{GeV}}
    \frac{\trh^3}{M_P^3}\left[1-\left(\frac{\rhorh}{\rhoe}\right)^{\frac{5}{6}}\right]^{-2} ,
    \label{ok}
\eeq
where $g_0 = 43/11$ and we use the Standard Model value $g_{\rm RH} = 427/4$.

We observe that $\Omega_X^{T, \, \xi} \propto \beta_1^{\xi} \, T_{\rm{RH}}^3$. Therefore large values of the couplings $\xi_{h}$ and $\xi_{X}$ would require a decrease in the reheating temperature. In Section \ref{Sec:results} we compare the scattering rates and the dark matter abundances with the minimally coupled case.

\subsection{$\phi~\phi \rightarrow X~X$}

Another mode of dark matter production is through the scattering of the inflaton itself. Whereas the graviton exchange channel 
was treated with care in \cite{MO, CMOV}, in the case of
non-minimal coupling
it suffices to replace $R_X^{T, \, \xi}$ in Eq.~(\ref{Eq:boltzmann3}) with the production rate~(\ref{eq:rateinfdm1}),
\beq
\frac{dY_X^{\xi}}{da}=\frac{\sqrt{3}M_P}{\sqrt{\rhorh}}a^2\left(\frac{a}{a_{\rm RH}}\right)^{\frac{3}{2}}R_X^{\phi, \, \xi}(a) \, , 
\label{Eq:boltzmann4}
\eeq
and to integrate 
between $a_{\rm end}$ and $a_{\rm RH}$, which leads to
\beq
n_X^{\phi, \, \xi} (a_{\rm RH}) = \frac{\sigma_{\phi X}^{\xi~2} \rho_{\rm{RH}}^{3/2}M_P}{4 \sqrt{3} \pi m_{\phi}^4} \left[\left( \frac{a_{\rm{RH}}}{a_{\rm{end}}}\right)^{\frac{3}{2}} - 1 \right]\sqrt{1 - \frac{m_{X}^2}{m_{\phi}^2}} \, .
\label{Eq:nsphi}
\eeq
For $a_{\rm RH} \gg a_{\rm end}$, using Eq.~(\ref{rhophia}) 
we can express $n_X^{\phi,\,\xi}$ as a function of $\rhoe$:
\beq
n_X^{\phi, \, \xi} (a_{\rm RH})\simeq \frac{\sigma_{\phi X}^{\xi~2} \rhorh \sqrt{\rhoe} M_P}{4 \sqrt{3} \pi m_{\phi}^4} \sqrt{1 - \frac{m_{X}^2}{m_{\phi}^2}} \, ,
\label{n0phi}
\eeq
and we find
\begin{flalign}
\frac{\Omega_X^{\phi, \, \xi} h^2}{0.12}\simeq \frac{1.3 \times 10^{7} \sigma_{\phi X}^{\xi~2}  \rho_{\rm{RH}}^{1/4} M_P^2}{m_{\phi}^3} \frac{m_X}{1 \, \rm{GeV}} \sqrt{1 - \frac{m_{X}^2}{m_{\phi}^2}} \, ,
\label{Eq:omegaphixi}
\end{flalign}
where we assumed the Starobinsky value for $\rhoe$. 
The analogous expression for models with minimally coupled gravity is found by replacing $\sigma_{\phi X}^{\xi} \rightarrow \sigma_{\phi X}$.

Up to this point we have assumed that the radiation is produced via the direct inflaton decay to a fermion pair. In the next subsection we discuss an unavoidable radiation production channel when the inflaton condensate scattering produces Higgs bosons in models with minimal and non-minimal coupling to gravity.

\subsection{$\phi~\phi \rightarrow h~h$}

Gravitational processes that produce dark matter can also 
populate the
thermal bath in the same way. Even if this 
Planck-suppressed production mechanism does not dominate throughout the entire reheating process, 
it was shown in
\cite{CMOV} that for $\trh \lesssim 10^{9}$ GeV  it is graviton 
exchange that dominates the production of the thermal bath at the
very beginning of the reheating, when $\rho_\phi\sim \rhoe$. 
In fact, it was shown that the maximal temperature reached, $\tmax$, 
(which can be considered as an absolute lower bound on $\tmax$)
is $\tmax\sim 10^{12}$ GeV.
It is therefore natural to determine the value of the couplings ($\xi_\phi$, $\xi_h$), for which
non-minimal gravitational processes generate the thermal bath at early times, and the 
maximal temperature which can be attained by these processes.

Following the discussion in the previous subsection,
to compute the radiation energy density produced by gravitational couplings 
we implement the rate 
$R_h^{\phi,\,\xi}$ (\ref{eq:ratehiggs}) into the Friedmann equation~(\ref{eq:diffrhor}) 
\begin{equation}
    \label{eq:rhorad1}
    \frac{d \rho_R}{dt} + 4H\rho_R \; \simeq \; N_h \frac{\sigma_{\phi h}^{\xi~2}}{8\pi} \frac{ \rho_{\phi}^2}{m_{\phi}^3} \, ,
\end{equation}
where we took into account that each scattering corresponds
to an energy transfer of $2 m_\phi$.\footnote{Or equivalently that each Higgs quanta 
carries an energy $m_\phi$.}
The solution to this equation is 
\begin{equation}
    \rho_R \; = \; N_h \frac{\sqrt{3} \sigma_{\phi h}^{\xi~2}}{4\pi} \frac{\rhoe^{3/2}M_P}{m_{\phi}^3} \left[ \left( \frac{a_{\rm{end}}}{a}\right)^4 - \left(\frac{a_{\rm{end}}}{a}\right)^{\frac{9}{2}} \right] \, .
    \label{rhoRphi}
\end{equation}
Note that the dependence on the scale factor $a$ is very different from 
that found in Eq.~(\ref{Eq:rhoR}) due to inflaton decay.
Indeed, the Higgs bosons produced by gravitational
scattering
(minimal as well as non-minimal) are redshifted to a greater extent because of the high dependence
of the rate on their energy due to the form of the energy-momentum
tensor $T_{\mu \nu}^0$. Since $\rho_R \propto a^{-4}$ in Eq.~(\ref{rhoRphi}) (at large $a$) and $\rho_\phi \propto a^{-3}$ in Eq.~(\ref{rhophia}), reheating through this process does not occur (i.e., $\rho_R$ never comes to dominate the total energy at late times) and inflaton decay is necessary.\footnote{This conclusion is avoided if the inflaton potential about minimum is approximated by $\phi^k$ with a higher power of $k>4$ \cite{CMOV,Haque:2022kez}.} 

However, as in the case of the reheating from the inflaton decay, 
the energy density in Eq.~(\ref{rhoRphi})
exhibits a maximum when $a = a_{\rm max} = (81/64) a_{\rm end}$. 
The maximum radiation density is then,
\begin{equation}
    \rho_{\rm{max}}^{\xi} \; \simeq \; N_h \frac{\sigma_{\phi h}^{\xi~2}}{12 \sqrt{3} \pi} \frac{\rhoe^{3/2}M_P}{m_{\phi}^3} \left(\frac{8}{9} \right)^8 \, ,
\end{equation}
and from this expression we find that the maximum temperature produced by gravitational interactions is given by
\bea
    T_{\rm{max}}^{\xi} &\simeq& 6.5 \times 10^{11} \left(\frac{|\sigma_{\phi h}^{\xi}|}{10^{-11}} \right)^{\frac{1}{2}} 
{\rm GeV} 
\label{tximax}
\\
&\simeq&1.8 \times 10^{12}\sqrt{|\xi|}\left(|5+12 \xi|\right)^{\frac{1}{2}} \left(\frac{m_{\phi}}{3\times10^{13}\,{\rm GeV}}\right)
 {\rm GeV}\, ,
 \nonumber
\eea
where we took $\xi_\phi=\xi_h=\xi$ in the last equality.
The analogous expression for models with minimally coupled gravity is found by replacing $\sigma_{\phi h}^{\xi} \rightarrow \sigma_{\phi h}$.

\begin{figure}[ht!] 
    \includegraphics[width = 1.0\columnwidth]{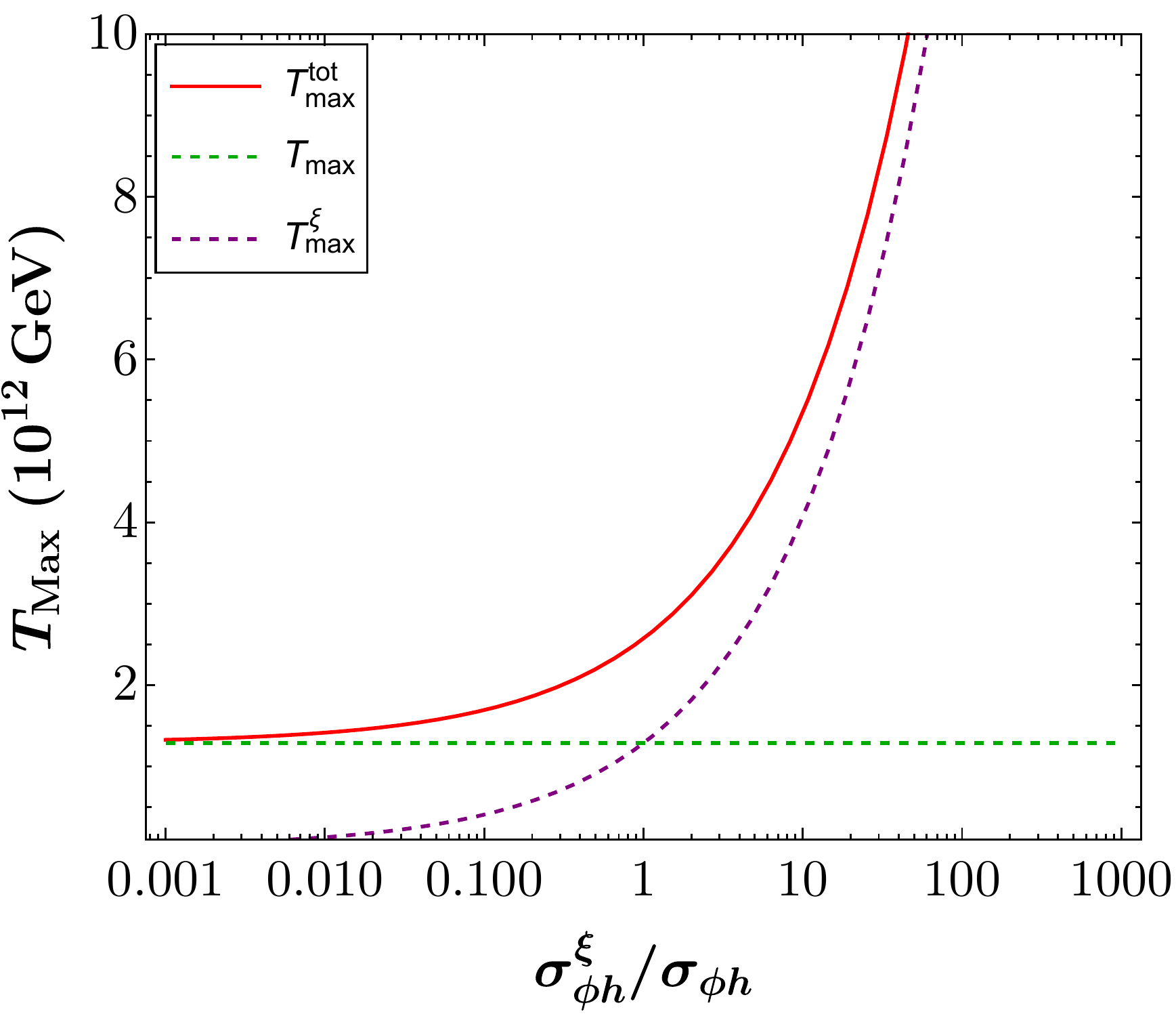}
\caption{\em \small The maximum temperature during reheating generated separately by minimal and non-minimal gravitational scattering of Higgs bosons in the thermal bath. }
    \label{plotTmaxxi}
\end{figure}

To compare the maximum temperature obtained by non-minimal interactions with respect to minimal gravitational interactions, we can rewrite Eq.~(\ref{tximax}) (now including minimal interactions in $\tmax^\xi$) as
\beq
   T_{\rm{max}}^{\xi} \simeq 1.3 \times 10^{12} \left(\frac{|\sigma_{\phi h}^{\xi}+\sigma_{\phi h}|}{\sigma_{\phi h}} \right)^{\frac{1}{2}} {\rm GeV} \, .
   \label{Eq:tximaxbis}
\eeq
The value of $\xi$ for which the maximum temperature
generated by the non-minimal coupling surpasses the one from graviton
exchange is shown in Fig.~\ref{plotTmaxxi} and is determined using
\beq
\sqrt{\frac{|\sigma_{\phi h}^\xi|}{|\sigma_{\phi h}|}}
=\sqrt{2 |\xi|}\left(|5 +12 \xi|\right)^{\frac{1}{2}} >1
\label{Eq:ratiotmax}
\eeq
which is satisfied when $\xi> 1/12$ or $\xi < -1/2$, as discussed earlier. 

As noted above and discussed in \cite{CMOV}, minimal (and non-minimal) gravitational interactions for a quadratic inflaton potential do not lead to the completion of the  reheating process, thus requiring additional inflaton interactions for decay. Although radiation density produced in scattering falls off faster than that from decay,
at early time, the radiation density may in fact dominate and determine $\tmax$.
To determine when the $\phi~\phi \rightarrow h~h$ process leads to the maximum temperature, we rewrite Eq.~(\ref{Eq:rhoR}) as:
\begin{equation}
\rho^y_R \;=\; \frac{\sqrt{3} y^2 m_{\phi} M_P^3}{20 \pi} \left(\frac{\rhoe}{M_P^4} \right)^{\frac{1}{2}} \left[
\left(\frac{\ae}{a}\right)^{\frac{3}{2}} - \left(\frac{\ae}{a}\right)^4 \right] \, .
\label{Eq:rhorsigma_sol2}
\end{equation}
Using Eq.~(\ref{Eq:amax}), we find that the maximum radiation density 
produced by the inflaton decay is given by
\begin{equation}
    \rho_{\rm max}^y \; = \; \frac{\sqrt{3} y^2 m_{\phi} M_P^3}{32 \pi} \left(\frac{\rhoe}{M_P^4} \right)^{\frac{1}{2}} \left(\frac{3}{8} \right)^{\frac{3}{5}} \, .
\end{equation}
The maximum temperature  is therefore determined by (non-minimal) gravitational 
interactions when 
\begin{equation}
    y^2 \lesssim N_h \frac{8 \rhoe \sigma_{\phi h}^{\xi~2}}{9 m_{\phi}^4} \left(\frac{8}{9} \right)^8 \left(\frac{8}{3} \right)^{\frac{3}{5}}
\end{equation}
or
\beq
y\lesssim 1.6~\sigma_{\phi h}^{\xi}
\sqrt{\frac{\rhoe}{m_\phi^4}}
\simeq 
5.4 \times 10^{4}~\sigma_{\phi h}^{\xi}  \left(\frac{3 \times 10^{13} \, \rm{GeV}}{m_{\phi}} \right) \, .
\label{Eq:ymin}
\eeq
This leads to the following reheating temperature:
\bea
    T_{\rm{RH}} \; &\lesssim& \; 3.1 \times 10^{19} \sigma_{\phi h}^{\xi}  \left(\frac{m_{\phi}}{3 \times 10^{13} \, \rm{GeV}} \right)^{-1/2}
    {\rm GeV} \nonumber
    \\
    &\lesssim& 2.4\times 10^9 \left(\frac{m_\phi}{3 \times 10^{13}} \right)^{\frac{3}{2}}\xi(5 + 12 \xi)~{\rm GeV}
\eea
where $\trh$ is given by \cite{GKMO2}
\beq
\rho_\phi(\arh)=\alpha \trh^4=
\frac{12}{25}\Gamma_\phi^2M_P^2=
\frac{3 y^4 m_{\phi}^2 M_P^2}{400 \pi^2} \, ,
\label{rhoRH}
\eeq
when the reheating temperature is determined by inflaton decay.

\begin{figure}[ht!] 
    \includegraphics[width = 1.0\columnwidth]{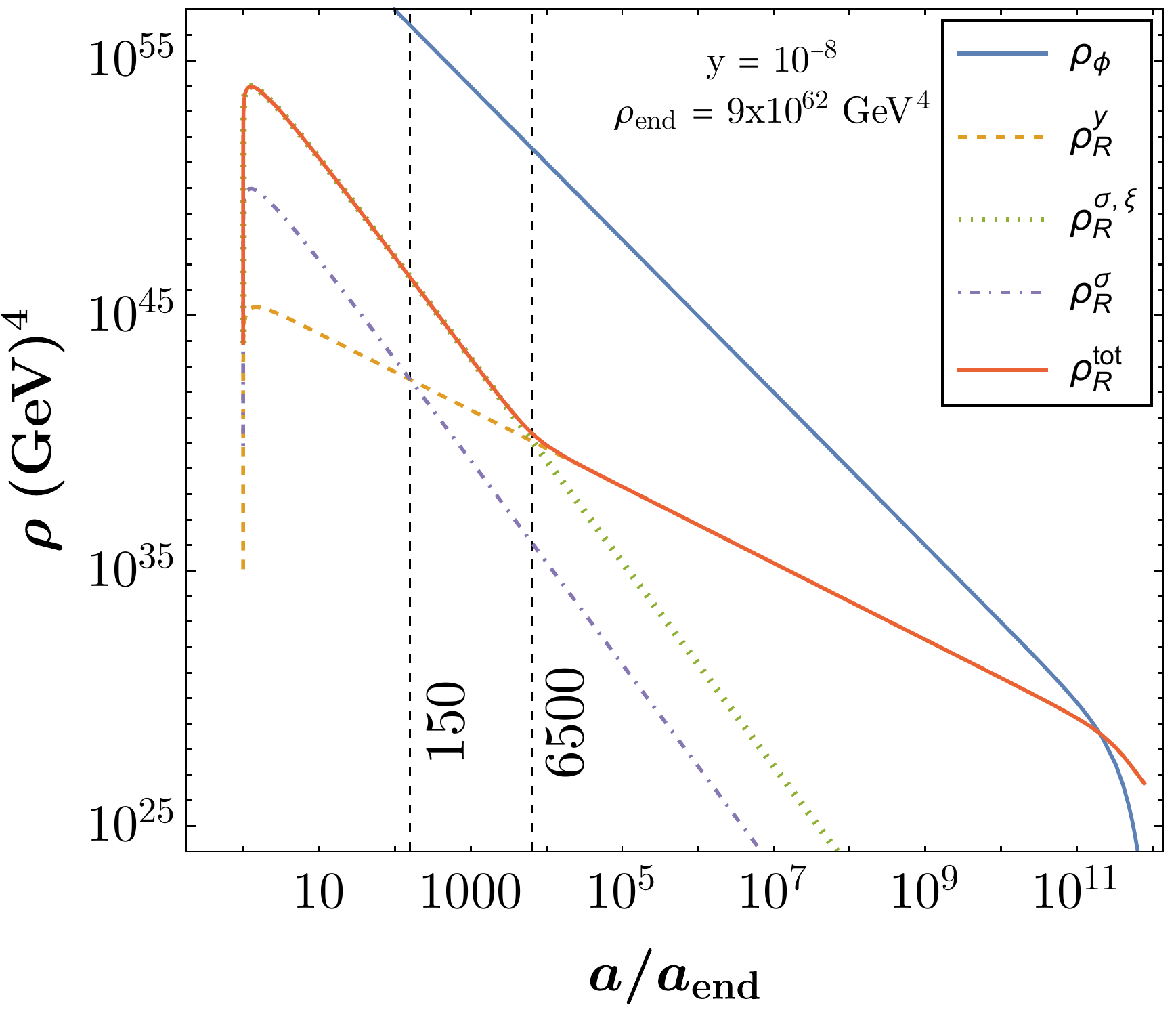}
\caption{\em \small Evolution of the inflaton density (blue) and the total radiation density (red), with radiation density produced from inflaton decays (dashed orange) and $\phi~\phi \rightarrow h~h$ scattering processes $\rho_R^{\sigma, \, \xi}$ (dotted green) and $\rho_R^{\sigma}$ (dash-dotted purple) with $\sigma_{\phi h}^{\xi}/\sigma_{\phi h} = 100$ (or $\xi_{\phi} = \xi_h = \xi \simeq -2.3~{or}~1.8$), as a function of $a/\ae$ for a Yukawa-like coupling $y = 10^{-8}$ and $\rhoe \simeq 0.175 \, m_{\phi}^2 M_P^2$ $\simeq 9 \times 10^{62} \, \rm{GeV}^4$. The black dashed lines corresponds to the ratios $a_{\rm{int}}/a_{\rm{end}} \simeq 150$ and $6500$, which agrees with Eq.~(\ref{aint}). The numerical solutions are obtained from Eqs.~(\ref{eq:diffrhophi}), (\ref{eq:diffrhor}), and (\ref{eq:rhorad1}).}
    \label{enden1}
\end{figure}

The  primary effect of the gravitational scattering processes on reheating is the augmentation of $\tmax$ for sufficiently small 
inflaton decay coupling, $y$. This can be seen 
in Fig.~\ref{enden1} where we show the evolution of the energy density of radiation from scattering and decay as well as the energy density of the inflaton as a function of $a/\ae$ for $\sigma_{\phi h}^{\xi} = 0$ and $\sigma_{\phi h}^{\xi}/\sigma_{\phi h} = 100$, respectively.

As we saw in Eq.~(\ref{Eq:ratiotmax}), minimal gravitational interactions dominate over non-minimal interactions when $\sigma_{\phi h}^{\xi} < \sigma_{\phi h}$ or when
\beq
12 \xi_\phi \xi_h + 3 \xi_h + 2 \xi_\phi < \frac12 \, ,
\label{xicond1}
\eeq
when we neglect contributions proportional to the Higgs mass. In this case, the maximum temperature
is determined by gravitational interactions when $y \lesssim 2.1 \times 10^{-6}$ from Eq.~(\ref{Eq:ymin}) using $\sigma_{\phi h}$ from Eq.~(\ref{sph}). The evolution of the energy densities in this case is shown in Fig.~\ref{enden1} with $y = 10^{-8}$. 
However as the energy density of radiation 
after the maximum falls faster than $\rho_\phi$,
reheating in the Universe is determined by the inflaton decay.
For a sufficiently small coupling $y$, the energy density from 
the decay dominates the radiation density at $a > a_{\rm int}$, where
\begin{equation}
    \frac{a_{\rm int}}{\ae} \simeq \left(\frac{5 \sigma_{\phi h}^2 N_h \rhoe}{y^2 m_\phi^4} \right)^{2/5} \simeq 1.6  \left( \frac{\sigma_{\phi h} M_P}{y m_\phi} \right)^{4/5}\, .
    \label{aint}
\end{equation}
For $\sigma_{\phi h} = 3.8 \times 10^{-11}$,
$m_\phi = 3 \times 10^{13}$ GeV, and $y = 10^{-8}$ we have $a_{\rm int} \approx 160 \ae$,
as seen in the figure. 

When Eq.~(\ref{xicond1}) is not satisfied, 
non-minimal interactions may dominate as shown in the bottom panel of Fig.~\ref{enden1}, for $\sigma_{\phi h}^{\xi}=100 \sigma_{\phi h}$ and $y = 10^{-8}$. The cross-over can be determined from Eq.~(\ref{aint}) with the replacement $\sigma_{\phi h} \to \sigma_{\phi h}^{\xi}$.
In this example, $a_{\rm int} \approx 6500 \ae$.

\section{Results}
\label{Sec:results}

We now turn to some general results that may be obtained from the framework described above.
Concerning the gravitational production of dark matter from 
the thermal bath, the difficulty of populating the Universe 
 via the exchange of a graviton was already known \cite{Bernal:2018qlk,CMOV}. 
Summing the minimal and non-minimal contributions in 
Eq.~(\ref{oh2Txi}), we find for $\rhorh\ll\rhoe$
\bea
\frac{\Omega_X^T}{0.12}&\simeq&
\left[1+30f(\xi_h,\xi_X)\right]
\left(\frac{\trh}{10^{14}~{\rm GeV}}\right)^3\left(\frac{m_X}{4.0 \times 10^9~ {\rm GeV}}\right)
\label{Eq:omegaTtot}
\nonumber
\\
&=&\left[1+120\xi^2(1+6 \xi+12\xi^2) \right] 
\nonumber  \\ 
&& \times 
\left(\frac{\trh}{10^{14}~{\rm GeV}}\right)^3\left(\frac{m_X}{4.0 \times 10^9~{\rm GeV}}\right)
\eea
with
\beq
f(\xi_h,\xi_X)=\xi_h^2 + 2\xi_h \xi_X + \xi_X^2 + 12\xi_h \xi_X \left(\xi_h + \xi_X + 4\xi_h \xi_X \right)
\nonumber
\eeq
where we assumed $\xi_h=\xi_X=\xi$ in the last equality, 
for simplicity. 
It is clear that, if we set $\xi=0$, i.e. if we consider
only graviton exchange, the reheating temperature necessary 
to obtain a reasonable density respecting the data \cite{Planck}
is dangerously close to 
the mass of the inflaton, even
for extremely large dark matter masses. 
This problem had already been raised in \cite{Bernal:2018qlk}
and resolved in \cite{MO,CMOV} by considering the dark matter produced from
the (minimal) gravitational inflaton scattering.

\begin{figure}[ht!] 
    \includegraphics[width = 1.0\columnwidth]{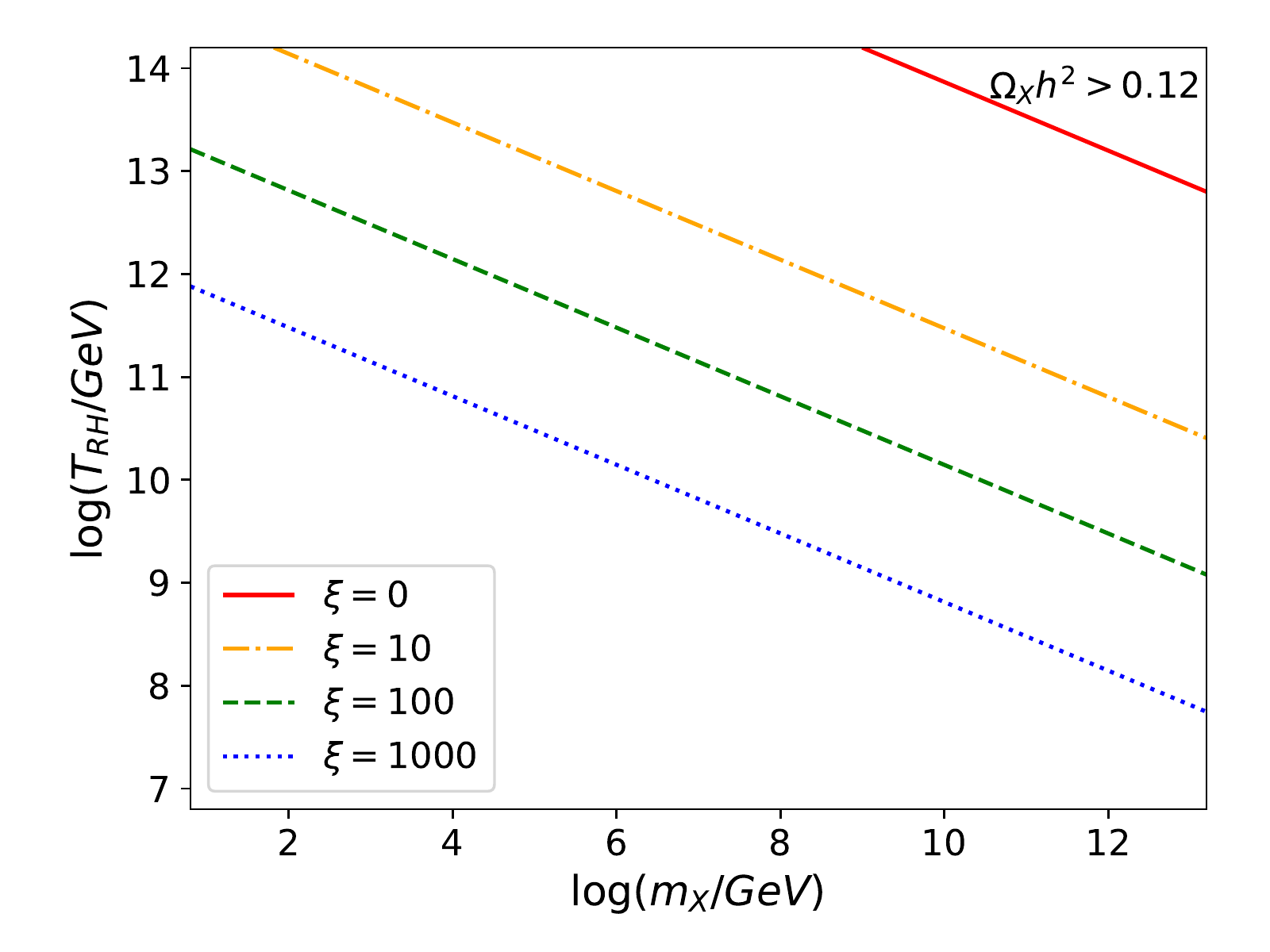}
\caption{\em \small Region of parameter
space respecting the relic density constraint 
$\Omega_X h^2=0.12$ in the plane ($m_X$,$\trh$) for different values of $\xi=\xi_h=\xi_X$ and 
$\rhoe \simeq 0.175 \, m_{\phi}^2 M_P^2$ in the
case of gravitational production from the thermal bath
$h~h \rightarrow X~X$. Both minimal and non-minimal 
contributions are taken into account.
}
    \label{Fig:omegaT}
\end{figure}

On the other hand, from Eq.~(\ref{Eq:omegaTtot}) 
we see that there is another solution to this tension if one allows 
for non-minimal gravitational couplings. Indeed, it is easy to see that for values of
$\xi_i \gtrsim 0.1$ ($f(\xi_h,\xi_X)\gtrsim{\frac{1}{30}}$), non-minimal gravitational production
dominates over graviton exchange. In this case,  it becomes easier to obtain the
correct dark matter density for more reasonable values of 
$\trh$ and/or $m_X$. For example, for a common value $\xi=\xi_h=\xi_X=1$, 
a temperature of $\trh\sim 1.2\times 10^{13}$ GeV, thus
slightly below the inflaton mass, is sufficient 
to produce an EeV dark matter candidate, whereas for 
$\xi=1000$,  $\trh \sim 10^{11}$ GeV will saturate the relic density for a 2.6 TeV dark matter mass. We show this result in Fig.~\ref{Fig:omegaT} where we plot the reheating temperature
needed to satisfy the relic density constraint as
function of $m_X$ for different value of $\xi$. For each value of $\xi$, the relic density exceeds $\Omega_X h^2 = 0.12$ above the corresponding curve.
As one can see, the line for $\xi=0$ is in the upper corner of the figure at high values of $\trh$ and $m_X$ and these drop significantly at higher values of $\xi$.

As was shown in \cite{MO, CMOV}, another possibility to avoid the necessity of 
high reheating temperatures and/or dark matter masses is the production of matter from the oscillations within the inflaton condensate when the energy stored
in the condensate is much larger than the reheating temperature. A simple comparison between Eqs.~(\ref{oh2Txi}) and (\ref{Eq:omegaphixi}) shows that the 
production of dark matter via inflaton scattering when $\xi_i \ne 0$ generally dominates over the production of dark matter from the thermal bath:
\bea
\frac{\Omega_X^{\phi, \, \xi}}{\Omega_X^{T, \, \xi}}
&\simeq& 34\frac{(\sigma_{\phi X}^\xi)^2}{\beta_1^\xi}
\frac{M_P^5}{\trh^2 m_\phi^3} \nonumber
\\
&\simeq&
185 \frac{M_P m_\phi}{\trh^2}\frac{(5+12 \xi)^2}{1+6 \xi +12 \xi^2}
\gg 1 \, ,
\
\eea
where we took $\xi=\xi_\phi=\xi_h=\xi_X$ and 
$m_X \ll m_\phi$ in the last equality.
We are therefore able to state that the relic density of dark matter 
generated by the non-minimal gravitational scattering of the inflaton is 
always much more abundant than that produced by the thermal bath.

Dark matter production from inflaton scattering via minimal graviton exchange also dominates over minimal gravitational thermal production \cite{CMOV}. This state of affairs is anything but surprising. 
Indeed, the energy available in the inflaton condensate at the onset of
oscillations is much greater than that available in 
the thermal bath during the reheating process. 
As the scattering cross-sections are themselves highly dependent 
on the energies through the energy-momentum tensor, 
it is quite normal that inflaton scattering is the 
dominant process for both minimal and non-minimal gravitational couplings.

Since inflaton scattering dominates in both the minimal and non-minimal gravitational interactions we can compare the two. We obtain 
\beq
\frac{\Omega_X^{\phi, \, \xi}}{\Omega_X^\phi} = \frac{\sigma_{\phi X}^{\xi~2}}{\sigma_{\phi X}^{2}}
\simeq 4 \xi^2(5 +12 \xi)^2 \, ,
\eeq
and we see again that non-minimal interactions dominate when $\xi > 1/12$ or $< -1/2$.
 
\begin{figure}[ht!] 
    \includegraphics[width = 1.0\columnwidth]{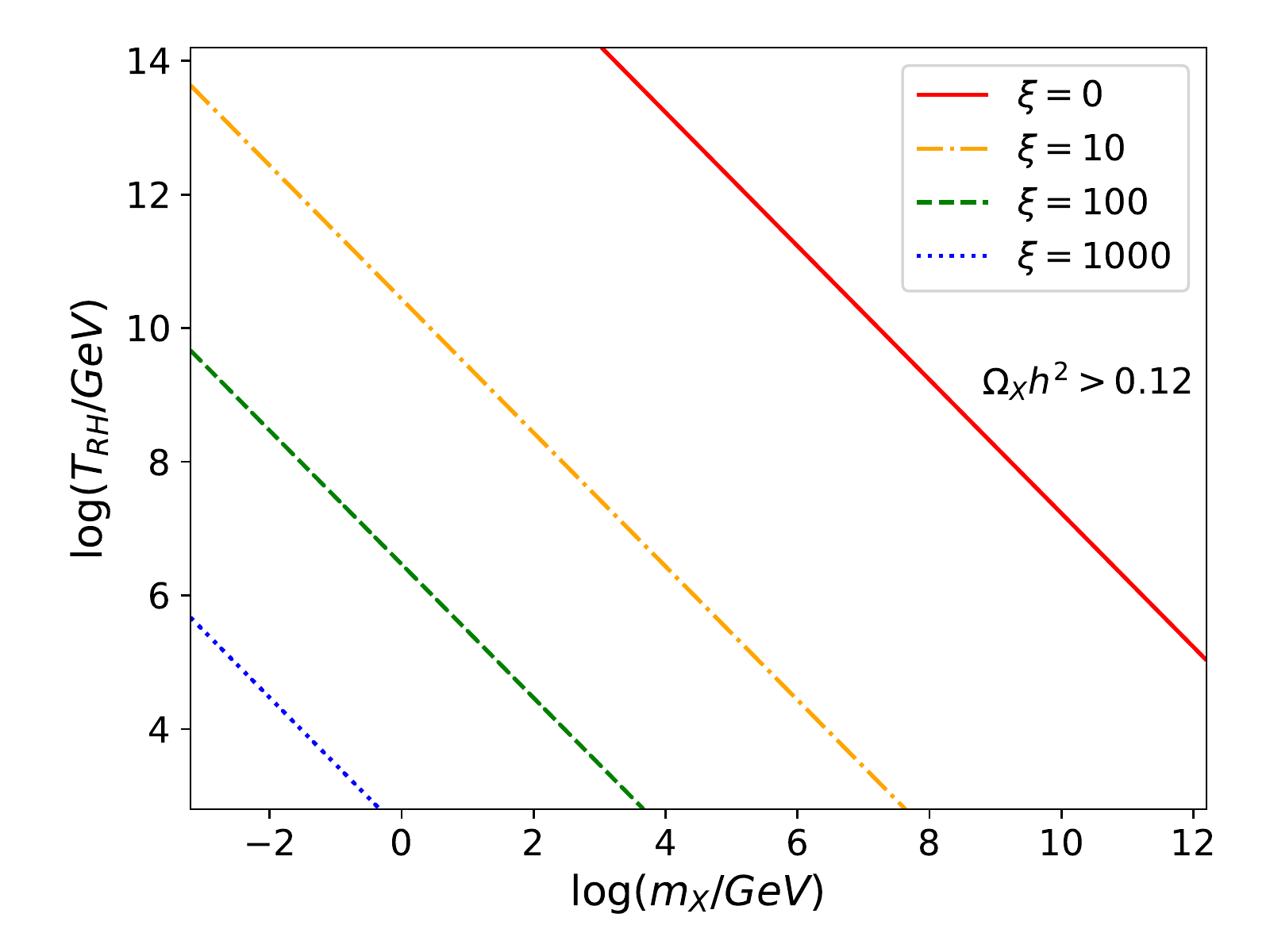}
\caption{\em \small Region of parameter
space respecting the relic density constraint 
$\Omega_X h^2=0.12$ in the plane ($m_X$,$\trh$) for different values of $\xi_\phi=\xi_h=\xi_X=\xi$ and 
$\rhoe \simeq 0.175 \, m_{\phi}^2 M_P^2$ in the
case of production from gravitational inflaton scattering
$\phi~\phi \rightarrow X~X$. Both minimal and non-minimal 
contributions are taken into account.}
    \label{Fig:omegatot}
\end{figure}

 We show in Fig.~\ref{Fig:omegatot} the region of the parameter space in the
 ($m_X$, $\trh$) plane allowed by the relic density constraint, adding all of the minimal and non minimal gravitational contributions, from inflaton scattering
 and as well as Higgs scattering from the thermal bath taking
 $\xi_\phi=\xi_h=\xi_X=\xi$. As expected, for 
 $\xi=0$ we recover the result found in \cite{CMOV}. As one can see, the difficulty in the
gravitational production from the thermal bath
 is indeed alleviated as a reheating temperature $\trh\simeq 10^{11}$ GeV
 allows for the production of a PeV scale dark matter candidate. If
 in addition we introduce the
 non-minimal couplings $\xi$, the necessary reheating temperature
to fit the Planck data may be as low as the electroweak scale 
for a GeV candidate if $\xi\gtrsim 1000$.

 Finally, we note that given the dark matter mass and
 reheating temperature (if that sector of beyond the Standard Model physics were known), the contours in Fig.~\ref{Fig:omegatot} allow us to place an upper bound on the non-minimal couplings, $\xi$.
 We can rewrite Eq.~(\ref{Eq:omegaphixi}) as
 \bea
\frac{\Omega_X h^2}{0.12} &=& 4.1\times 10^{-7} 
(12 \xi^2 + 5\xi +\frac12)^2
\left(\frac{\trh}{10^{10} {\rm GeV}} \right)
 \nonumber \\
&& \times \left(\frac{m_X}{1 {\rm GeV}} \right)  \left(\frac{m_\phi}{3\times 10^{13} {\rm GeV}} \right) \, ,
\eea
when $m_X \ll m_\phi$ and $\xi = \xi_\phi = \xi_X$.
Then, for example, if $m_X = 1$ TeV, and $\trh = 10^9$ GeV, we obtain an upper limit of $|\xi| \lesssim 4$.

\section{Conclusions}
\label{Sec:conclusion}

In this paper, we have generalized the 
{\it minimal} gravitational interactions
in the early Universe, i.e., the s-channel exchange of a graviton, to include {\it non-minimal} 
couplings of all scalars to the Ricci curvature $R$. We consider
a scalar sector $S_i$ consisting of the inflaton condensate $\phi$, the Higgs field $H$ and a dark matter candidate $X$, and
we have analyzed the impact of couplings of the type 
$\xi_i S_i^2 R$ on the reheating 
process and dark matter production. 
The latter can be generated
by the thermal Higgs 
scattering or excitations of the inflaton,
both through minimal and non-minimal gravitational couplings.
Whereas the Higgs scattering through the exchange 
of a graviton necessitates
a very large reheating temperature and/or dark matter mass in order to fulfill Planck CMB
constraints ($\trh \simeq 10^{14}$ GeV with $m_X \simeq 10^9$ GeV), for $\xi \gtrsim 0.1$,
the non-minimal coupling dominates the process and
alleviates the tension.
For $\xi \simeq 1000$, a dark matter mass of $\sim 1$ PeV with
$\trh\simeq 10^{10}$ GeV will satisfy the constraint, see Fig.~\ref{Fig:omegaT}.
However, thermal production is not the sole source of dark matter production through gravity. 
When we include the contribution (necessarily present) of the 
inflaton scattering, 
we showed that the energy stored in the condensate
at the end of inflation compensates largely the reduced
gravitational Planck coupling. These processes yield
the correct relic abundance through minimal graviton exchange 
for a dark matter mass of $\sim 10^8$ GeV
with $\trh\simeq10^{10}$ GeV, and the constraint is satisfied for a dark matter mass of  $\sim 100$ GeV
and $\trh \gtrsim 10^4$ GeV if one adds non-minimal 
couplings of the order
$\xi \simeq 100$ as we show in Fig.~\ref{Fig:omegatot}. 
Gravitational inflaton scattering also affects the reheating process, producing a maximum temperature 
$ \simeq 10^{12}$ GeV with minimal couplings, reaching as large as
$\tmax^\xi\simeq 5 |\xi| \tmax \simeq 10^{14}$ GeV for $\xi=100$ as one can see in Fig.~\ref{plotTmaxxi}. 
This result can be re-expressed as an upper limit to
$|\xi|$ given values of $m_X$ and $\trh$.

We can not over-emphasize 
that all of our results are unavoidable, in the sense that
they are purely gravitational, and do not rely on physics beyond the Strandard Mode. 
The relic density of dark matter, and maximum temperature of the thermal bath computed here should be considered as lower bounds, that should be implemented in any extension of the Standard Model, 
whatever is its nature.

 Note added : During the completion of the manuscript,
some overlapping results were presented in  \cite{Aoki:2022dzd}. 

\noindent {\bf Acknowledgements. }  The authors want 
to thank Emilian Dudas for useful discussions. This work was made possible by with the support of the Institut Pascal at 
Université Paris-Saclay during the 
Paris-Saclay Astroparticle Symposium 2021, with the support of the P2IO Laboratory of Excellence (program “Investisse ments d’avenir” ANR-11-IDEX-0003-01 Paris-Saclay and ANR-10-LABX-0038), the P2I axis of the Graduate School Physics of Université Paris-Saclay, as well as IJCLab, CEA, IPhT, APPEC, the IN2P3 master projet UCMN and EuCAPT
ANR-11-IDEX-0003-01 Paris-Saclay and ANR-10-LABX-0038). This project has received support from the European Union’s Horizon 2020 research and innovation programme under the Marie Sk$\lslash$odowska-Curie grant agreement No 860881-HIDDeN.
The work of K.A.O. and A.S.~was supported in part by DOE grant DE-SC0011842  at the University of
Minnesota.

\section*{Appendix}
\appendix
\renewcommand{\thesubsection}{\Alph{subsection}}
\subsection{PARTICLE PRODUCTION WITH A NON-MINIMAL COUPLING}
\label{sec:appendixA}
 
The full Jordan frame action we consider is given by Eq.~(\ref{eq:jordan}). 
The conformal transformation to the Einstein frame is given by
\begin{equation}
    g_{\mu \nu} \; = \; \Omega^2 \tilde{g}_{\mu \nu} \, ,
    \label{appa:conformal}
\end{equation}
where $g_{\mu \nu}$ is the Einstein frame spacetime metric and the conformal factor is expressed by Eq.~(\ref{eq:conformalfact}). It can readily be shown that the scalar curvature transforms as (see, e.g., \cite{Fujii:2003pa})
\begin{equation}
    \tilde{R} = \Omega^2 \left[R + 6 g^{\mu \nu} \nabla_{\mu} \nabla_{\nu} \ln \Omega - 6 g^{\mu \nu}\left(\nabla_{\mu} \ln \Omega \right)\left(\nabla_{\nu} \ln \Omega  \right)\right] \, .
\end{equation}
After eliminating the total divergence term, we find the Einstein frame action~(\ref{eq:einstein}).

To find the effective interaction terms we assume the small field limit~(\ref{eq:smallfield}) and expand the conformal factors in the Einstein frame action. We find the following effective interaction Lagrangian:
\begin{widetext}
\begin{eqnarray}
    \mathcal{L}_{\rm{eff}} \; &=& \; -\frac{1}{2} \left(\frac{\xi_{\phi} \phi^2}{M_P^2} +\frac{\xi_{X} X^2}{M_P^2} \right) \partial^{\mu} h \partial_{\mu} h -\frac{1}{2} \left(\frac{\xi_{h} h^2}{M_P^2} +\frac{\xi_{X} X^2}{M_P^2} \right) \partial^{\mu} \phi \partial_{\mu} \phi -\frac{1}{2} \left(\frac{\xi_{\phi} \phi^2}{M_P^2} +\frac{\xi_{h} h^2}{M_P^2} \right) \partial^{\mu} X \partial_{\mu} X \nonumber \\
    &+& \frac{6 \xi_{h} \xi_{X} h X}{M_P^2} \partial^{\mu} h \partial_{\mu} X + \frac{6 \xi_{h} \xi_{\phi} h \phi}{M_P^2} \partial^{\mu} h \partial_{\mu} \phi + \frac{6 \xi_{\phi} \xi_{X} \phi X}{M_P^2} \partial^{\mu} \phi \partial_{\mu} X + m_X^2 X^2 \left(\frac{\xi_{\phi} \phi^2}{M_P^2} +\frac{\xi_{h} h^2}{M_P^2} \right) \nonumber \\
    &+& m_\phi^2 \phi^2 M_P^2 \left(\frac{\xi_X X^2}{M_P^2} + \frac{\xi_h h^2}{M_P^2} \right) + m_h^2 h^2 \left(\frac{\xi_{\phi} \phi^2}{M_P^2} + \frac{\xi_X X^2}{M_P^2} \right) \, ,
\end{eqnarray}
\end{widetext}
and we can rewrite the above Lagrangian in terms of the effective couplings as Eq.~(\ref{lag4point}), with
\begin{eqnarray}
    \label{appa:sighx}
    \sigma_{h X}^{\xi} \; &=& \; \frac{1}{4M_P^2} \left[ \xi_h(2m_X^2 +s) +\xi_X  ( 2m_h^2 + s) \right. \nonumber \\
    &+& \left.  \left(12 \xi_X \xi_h(m_h^2 + m_X^2 - t) 
     \right) 
      \right] \, ,
\end{eqnarray}
\begin{equation}
    \label{appa:sigphiX}
    \sigma_{\phi X}^{\xi} \; = \; \frac{1}{2M_P^2} \left[\xi_{\phi} m_X^2 + 12 \xi_{\phi} \xi_{X} m_{\phi}^2 + 3 \xi_X m_{\phi}^2 + 2 \xi_{\phi} m_{\phi}^2 \right] \, ,
\end{equation}
\begin{equation}
    \label{appa:sigphih}
    \sigma_{\phi h}^{\xi} \; = \; \frac{1}{2M_P^2} \left[\xi_{\phi} m_h^2 + 12 \xi_{\phi} \xi_h m_{\phi}^2 + 3 \xi_h m_{\phi}^2 + 2 \xi_{\phi} m_{\phi}^2 \right] \, ,
\end{equation}
where $s, t$ are the Mandelstam variables.
The latter couplings assume an inflaton condensate in the initial state rather than a thermal Higgs in the initial state accounting for the lack of symmetry in the three couplings. 

\subsection{THERMAL PRODUCTION}
\label{sec:appendixB}
In this appendix we calculate the thermal dark matter production rate $R_X^{T, \, \xi}(T)$ arising from the effective four-point interaction $\sigma_{h X} h^2 X^2$, where $\sigma_{h X}$ is given by Eq.~(\ref{appa:sighx}). We also calculate the production rate $R_X^T(T)$ for the thermal scattering processes mediated by gravity alone, ${\rm{SM}}~{\rm{SM}} \rightarrow X~X$, 
that are unavoidable in models with a minimal coupling to gravity ($\xi_{\phi, h, X} = 0$)~\cite{Bernal:2018qlk, CMOV}, and compare the two results.

The production rate $R_X^{T, \, \xi}(T)$ can be computed from Eq.~(\ref{eq:rategen}). The matrix element squared is given by
\begin{equation}
    |\overline{{\cal M}}^{h X, \, \xi}|^2 \; = \; 4 \sigma_{hX}^{\xi~2} \, ,
\end{equation}
where in the limit where the Higgs boson mass is neglected, the Mandelstam variables $s$ and $t$ are given by
\begin{align}
    t\,= & \, \dfrac{s}{2}\left( \sqrt{1 - \frac{4m_X^2}{s}} \cos \theta_{13}-1 \right) + m_X^2 \, ,  \\  s\, = & \,2E_1E_2 \left(1-\cos \theta_{12} \right) \, .
\end{align}
We find the following coefficients for Eq.~(\ref{eq:ratefull1})
\begin{align}
    \label{eq:b1xi}
    \beta_1^{\xi} &\; = \; \frac{\pi^3}{2700} \left[\xi_h^2 + 2\xi_h \xi_X + \xi_X^2 + 12\xi_h \xi_X \left(\xi_h + \xi_X + 4\xi_h \xi_X \right)\right] \, , \\
    \label{eq:b2xi}
    \beta_2^{\xi} &\; = \; \frac{\zeta(3)^2 \xi_h}{2 \pi^5} \left[\xi_h + \xi_X + 6\xi_h \xi_X - 12 \xi_h \xi_X^2 \right] \, , \\
    \label{eq:b3xi}
    \beta_3^{\xi} &\; = \; \frac{\xi_h^2}{576 \pi} \, .
\end{align}
Similarly, using Eqs.~(\ref{eq:partialamp})-(\ref{eq:partialamp2}), we find the matrix element squared for minimally coupled gravity:
\begin{equation}
     |\overline{{\cal M}}^{h X}|^2 \; = \; \frac{1}{4M_P^4} \frac{\left(t(s+t) - 2m_X^2t + m_X^4 \right)^2}{s^2},
\end{equation}
where we have neglected the Higgs field mass. We find the coefficients:
\begin{align}
    \label{eq:b1}
    \beta_1 &\; = \; \frac{\pi^3}{81 000} \, , \\
    \label{eq:b2}
    \beta_2 &\; = \; - \frac{\zeta(3)^2}{30 \pi^5} \, , \\
    \label{eq:b3}
    \beta_3 &\; = \; \frac{1}{4320 \pi} \, .
\end{align}
Note that when both contributions are kept, and we neglect $m_h \ll m_X$, 
the full coefficients (including interference) are given by
\bea
\label{eq:b11}
\beta_1^\xi & = & \frac{\pi^3}{81 000} \left[ 30 \xi_h^2 \left(12 \xi_X (4 \xi_X + 1) + 1\right) \right. \nonumber \\
\label{eq:b22}
&& \left. + 10 \xi_h ( 6\xi_X + 1)^2 + 10 \xi_X (3 \xi_X + 1) + 1 \right] \, , \\
 \beta_2^\xi & = &  - \frac{\zeta(3)^2}{60 \pi^5}\left[ 2 + 10\xi_X \right.  \nonumber \\
&& \left.  + 5\xi_h\left(1 + 6\xi_X + 6\xi_h\left(6\xi_X(2\xi_X -1) -1 \right)\right) \right] \, , \\
    \beta_3^\xi & = & \frac{1}{8640 \pi}\left[2 +5\xi_h\left(32\xi_h-2\right)\right] \, .
      \label{eq:b33}
\eea
which reduces to Eqs.~(\ref{eq:b1}-\ref{eq:b3}) when all $\xi_i = 0$.

\end{document}